# Towards nonlinear quantum Fokker-Planck equations

Roumen Tsekov

DWI, RWTH, 52056 Aachen, Germany

It is demonstrated how the equilibrium semiclassical approach of Coffey *et al.* can be improved to describe more correctly the evolution. As a result a new semiclassical Klein-Kramers equation for the Wigner function is derived, which remains quantum for a free quantum Brownian particle as well. It is transformed to a semiclassical Smoluchowski equation, which leads to our semiclassical generalization of the classical Einstein law of Brownian motion derived before. A possibility is discussed how to extend these semiclassical equations to nonlinear quantum Fokker-Planck equations based on the Fisher information.

It is well known that the classical Fokker-Planck equations are linear due to the logarithmic dependence of the Boltzmann entropy on the probability density. More complex systems, however, exchange non-Shannon information as well, which results in nonlinear Fokker-Planck equations (Chavanis 2004, Frank 2005). In many systems, including quantum ones, the Fisher information plays an important role and reflects the quality of measurements (Frieden 1998). The scope of the present paper is to generalize the linear semiclassical Klein-Kramers equation derived by Coffey *et al.* (2007) to a nonlinear quantum Fokker-Planck equation, accounting for the Fisher information $FI \equiv \int \rho (\partial_x \ln \rho)^2 dx = -\int \rho \partial_x^2 \ln \rho \, dx$ stored in the quantum probability density $\rho$ (Reginatto 1998); the Shannon information reads $SI \equiv -\int \rho \ln \rho \, dx$.

Last year Coffey *et al.* (2007) introduced heuristically the following semiclassical Klein-Kramers equation

$$\partial_t W + \frac{p}{m}\partial_x W - \partial_x V \partial_p W + \frac{\hbar^2}{24}\partial_x^3 V \partial_p^3 W = b\partial_p[\frac{p}{m}W + (k_B T + \frac{\hbar^2}{12mk_B T}\partial_x^2 V)\partial_p W] \qquad (1)$$

describing the evolution of the Wigner function $W(x,p,t)$ of a quantum Brownian particle. Here $V(x)$ is an arbitrary potential and $b$ is the Brownian particle friction coefficient. A novelty in Eq. (1) is the quantum term on the right hand side, which ensures that the equilibrium semiclassical Wigner distribution, known from the statistical thermodynamics, is also a solution of Eq. (1). The result of Coffey *et al.* is very important since it corrects the Caldeira-Leggett master equation, which is believed to be rigorous. Obviously this is not the case since the latter does not provide the exact quantum canonical Gibbs distribution at equilibrium. However, how it was demonstrated in (Tsekov 2007), Eq. (1) does not describe well enough the evolution since

in the case of a free quantum Brownian particle ($V = 0$) it reduces to the classical Klein-Kramers equation

$$\partial_t W + \frac{p}{m}\partial_x W = b\partial_p(\frac{p}{m}W + k_B T \partial_p W) \qquad (2)$$

Hence, according to Coffey *et al.* a free quantum Brownian particle behaves as a classical one, which is only true if it moves in vacuum ($b = 0$) since in this case no energy exchange takes place. The contemporary theory of decoherence also misinterprets Eq. (2). Usually, supposing an initially quantum Wigner function and studying then its evolution via Eq. (2), one concludes that the quantum nature reduces in time to the classical one (Zurek 2003). This is, however, inconsistent since no matter of the initial conditions the long-time solution of the classical Klein-Kramers equation is classical as well due to irreversibility. Hence, the rigorous way is to find a quantum generalization of Eq. (2), which is the goal of the present paper.

The problem in Eq. (1) appears from a non-rigorous extension of an equilibrium result to the non-equilibrium case. Thus, the temperature dependence of the quantum term on the right hand side of Eq. (1) clearly shows its equilibrium origin and validity. The identification of a dynamic equation from its equilibrium solution is not unique. Because we are looking for a semiclassical approximation, one expects that the temperature dependence of the quantum term on the right hand site of Eq. (1) originates from the equilibrium classical probability density. Since the latter is the Boltzmann distribution $\rho_{cl}^{eq} = \exp(-V/k_B T)/Z$, one could replace in the non-equilibrium case the term $\partial_x^2 V / k_B T = -\partial_x^2 \ln \rho_{cl}^{eq}$ of Eq. (1) by $-\partial_x^2 \ln \rho_{cl}$ to obtain an alternative semiclassical Klein-Kramers equation

$$\partial_t W + \frac{p}{m}\partial_x W - \partial_x V \partial_p W + \frac{\hbar^2}{24}\partial_x^3 V \partial_p^3 W = b\partial_p[\frac{p}{m}W + (k_B T - \frac{\hbar^2}{12m}\partial_x^2 \ln \rho_{cl})\partial_p W] \qquad (3)$$

which provides also the correct semiclassical equilibrium distribution. However, in contrast to Eq. (1), Eq. (3) keeps the new quantum term in the case of a free quantum Brownian particle ($V = 0$) as well

$$\partial_t W + \frac{p}{m}\partial_x W = b\partial_p[\frac{p}{m}W + (k_B T - \frac{\hbar^2}{12m}\partial_x^2 \ln \rho_{cl})\partial_p W] \qquad (4)$$

In the case of strong friction one can easily derive from Eq. (4) a semiclassical Smoluchowski equation for the probability density in the coordinate space $\rho = \int W dp$

$$\partial_t \rho = \partial_x^2 [(D - \frac{\hbar^2}{12mb} \partial_x^2 \ln \rho_{cl}) \rho] \tag{5}$$

where $D = k_B T / b$ is the Einstein diffusion constant. The classical solution of Eq. (5) is the well-known Gaussian distribution

$$\rho_{cl} = \frac{1}{\sqrt{4\pi Dt}} \exp(-\frac{x^2}{4Dt}) \tag{6}$$

Substituting Eq. (6) in Eq. (5) leads to a semiclassical diffusion equation

$$\partial_t \rho = D(1 + \lambda_T^2 / 6Dt) \partial_x^2 \rho \tag{7}$$

where $\lambda_T = \hbar / 2\sqrt{mk_B T}$ is the thermal de Broglie wavelength. The expression in the brackets represents the relative increase of the diffusion constant by quantum effects. The solution of Eq. (7) is also a Gaussian distribution density with dispersion $\sigma^2$ satisfying the following equation

$$\partial_t \sigma^2 = 2(D + \lambda_T^2 / 6t) \tag{8}$$

Integrating on time yields the already known semiclassical generalization of the classical Einstein law of Brownian motion

$$\sigma^2 = 2Dt + \lambda_T^2 \ln(6Dt / \lambda_T^2) / 3 \tag{9}$$

which is derived from a nonlinear quantum Smoluchowski equation (Tsekov 2007). Note that Eq. (9) is applicable for large times $t > \lambda_T^2 / 2D$ only, while at shorter times the diffusion is strongly quantum and cannot be described by the semiclassical approach (Tsekov 2009).

As is seen, Eq. (4) is linear only because of the semiclassical linearization of the quantum term but in general one expects the quantum Klein-Kramers equation to be nonlinear (Alicki and Messer 1983, Doebner and Goldin 1994, Tsekov 2009). As a first step towards this direction one could replace $\rho_{cl}$ by $\rho$ in Eq. (4) to obtain

$$\partial_t W + \frac{p}{m} \partial_x W = b \partial_p [\frac{p}{m} W + (k_B T - \frac{\hbar^2}{12m} \partial_x^2 \ln \rho) \partial_p W] \tag{10}$$

This nonlinear equation shows that the quantum effects correspond to effective increase of the temperature. It is, however, not surprising, when one realizes that the quantum term in Eq. (10) is in fact the so-called local quantum temperature (Sonego 1991) given by the expression

$$k_B T_Q = -\frac{\hbar^2}{4m} \partial_x^2 \ln \rho \tag{11}$$

The mean value of the quantum temperature is proportional to the Fisher information. For a Gaussian distribution the momentum dispersion $m k_B T_Q = \hbar^2 / 4\sigma^2$ corresponding to the quantum temperature is just the minimal Heisenberg uncertainty. A nonlinear quantum Smoluchowski equation (Ancona and Iafrate 1989) can be derived from Eq. (10)

$$\partial_t \rho = \partial_x^2 [(D + D_Q / 3)\rho] = \partial_x (D \partial_x \rho + \rho \partial_x Q / 3b) \tag{12}$$

where the quantum temperature is included in a quantum diffusion coefficient $D_Q = k_B T_Q / b$. In the last expression of Eq. (12) the Bohm quantum potential $Q = -\hbar^2 \partial_x^2 \sqrt{\rho} / 2m\sqrt{\rho}$ is introduced. Note that the mean value of $Q$ is also proportional to the Fisher information (Reginatto 1998). The solution of Eq. (12) is also a Gaussian distribution density with dispersion given by

$$\sigma^2 - \lambda_T^2 \ln(1 + 3\sigma^2 / \lambda_T^2) / 3 = 2Dt \tag{13}$$

Substituting $\sigma^2$ in the logarithmic term by the classical expression $2Dt$ yields an improved semiclassical limit $\sigma^2 = 2Dt + \lambda_T^2 \ln(1 + 6Dt / \lambda_T^2) / 3$, which is always positive and recovers Eq. (9) at large time. At short time Eq. (13) reduces to $\sigma^2 = \hbar\sqrt{t/3mb}$, which is close to the exact expression $\sigma^2 = \hbar\sqrt{t/mb}$ for the purely quantum diffusion (Tsekov 2009). This indicates, however, that Eq. (12) is approximate, since it is derived from semiclassical considerations. As is shown in (Tsekov 2007), the factor 3 originates from an integral over the reciprocal temperature $\beta = 1/k_B T$. A more general analysis of the thermo-quantum diffusion is given in (Tsekov 2009) and the enhanced nonlinear quantum Smoluchowski equation reads

$$\partial_t \rho = D \partial_x (\partial_x \rho + \rho \partial_x \int_0^\beta Q d\beta)_b \tag{14}$$

where the subscript $b$ indicates that the friction coefficient should be considered constant during the integration on $\beta$. The integral in Eq. (14) accounts for the interplay between the thermal and quantum motion. Hence, the total entropy $-\ln \rho - \beta Q + \int_0^\beta \beta dQ$ is not simply a super-position of Shannon and Fisher components. The last integral represents the thermo-quantum entropy (Tsekov 2009), which reflects the temperature dependence of the quantum potential.

The solution of Eq. (14) is a Gaussian distribution again with dispersion satisfying the following equation

$$\partial_t \sigma^2 = 2D(1+\sigma^2 \int_0^\beta \frac{\hbar^2}{4m\sigma^4} d\beta)_b \tag{15}$$

Using the auto-model expression $\sigma^2/\hbar\sqrt{t/mb} = y(\sqrt{mt/b}/\beta\hbar)$ Eq. (15) can be transformed into a nonlinear ordinary differential equation

$$yy'' - y'^2 - y^2/x^2 + 4y' + 1/x^2 = 0 \tag{16}$$

where $x = \sqrt{mt/b}/\beta\hbar = \sqrt{Dt}/2\lambda_T$. The boundary conditions relevant to Eq. (16) are $y(0)=1$ and $y'(\infty)=2$.

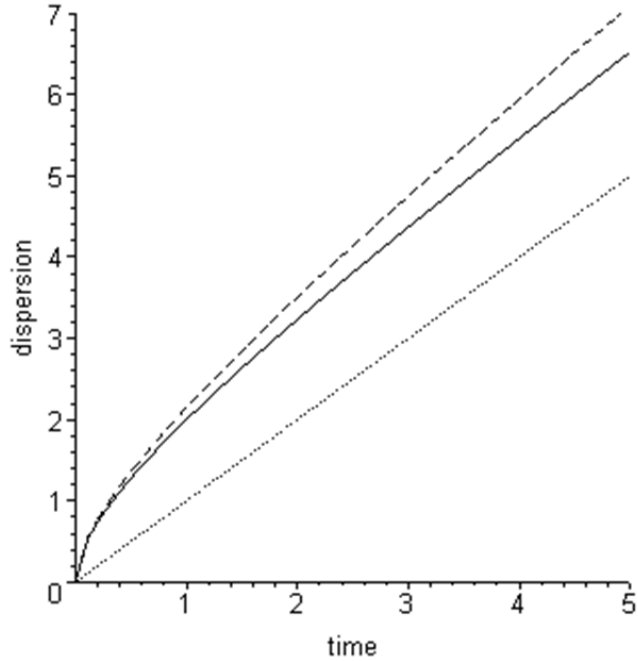

**Fig. 1** The universal dependence of the dispersion $\sigma^2/\lambda_T^2$ vs. time $2Dt/\lambda_T^2$ from the solution of Eq. (16) (solid line) compared to the approximation $\sigma^2/\lambda_T^2 - \ln(1+\sigma^2/\lambda_T^2) = 2Dt/\lambda_T^2$ (Tsekov 2009) (dashed line). The dotted line presents the classical Einstein law.

The numerical solution of Eq. (16) is plotted in Fig. 1 as well as the analytical approximation $\sigma^2/\lambda_T^2 = -1 - W_{-1}[-\exp(-1-2Dt/\lambda_T^2)]$, where $W_{-1}$ is a Lambert function. The latter is derived from Eq. (15) via neglecting of the thermo-quantum entropy (Tsekov 2007). The comparison confirms that the analytical expression is an upper limit of the exact solution (Tsekov 2009) and that the thermo-quantum entropy is not very essential for a free quantum Brownian particle. This is not the case of the Fisher information, however, which seems to be the essence of the quantum mechanics (Reginatto 1998). It is not surprising since the quality of measurements is of crucial importance in the quantum world. For instance, the Fisher information appears in atomic physics as the so-called von Weizsäcker (1935) correction of the Thomas-Fermi-Dirac density functional model.